# Buoyancy storms in a zonal stream on the polar beta-plane: experiments with altimetry

Y. Sui and Y. D. Afanasyev[a]

*Memorial University of Newfoundland, St. John's, Canada*

[a]Corresponding author address:

Yakov Afanasyev

Memorial University of Newfoundland, St. John's, NL, Canada

E-mail: afanai@mun.ca

http://www.physics.mun.ca/~yakov

**Abstract**

Results from a new series of experiments on flows generated by localized heating in the presence of a background zonal current on the polar β-plane are presented. The flow induced by a heater without the background zonal flow is in the form of a β-plume. Zonal jets of alternating directions are formed within the plume. The westward transport velocity in the plume is proportional to the upwelling velocity above the heater in agreement with linear theory. When the background flow in the form of the eastward zonal current is present, the β-plume can be overwhelmed by the eastward current. The main control parameters of the experiment are the strength of the heater and strength of the sink which is used to create the background flow. The regime diagram shows the area where a β-plume can exist in the parameter space. The critical value of the velocity of the zonal flow below which the β-plume can exist is obtained by considering barotropic Rossby waves emitted by the baroclinic eddies in the heated area.

## I. INTRODUCTION

Spatially localized perturbations which can be associated with sources or sinks often occur in atmospheres of planets and in the Earth's oceans. Zonal currents are also a typical feature of these environments where the variation of the Coriolis parameter (β-effect) is dynamically important. It is important therefore to study the interactions between the flows generated by the sink/source on the β-plane and the background zonal flow. Several examples of these flows, quite different in their nature yet having similar general dynamics, can be given here.

Hydrothermal vents in the ocean create upwelling of warm water. Warm water rising to the surface can be considered as an effective source at the surface (and as a sink at the bottom where the rising water entrains ambient cold water). Stommel[1] showed that this source-sink combination generates β-plumes in the absence of any background zonal flow. A β-plume is an essentially zonal circulation westward of the source/sink consisting of two currents/jets flowing in opposite directions and reconnected in the area where the sink or source is located. A source generates a westward jet at its southern flank and an eastward jet at its northern flank. The direction of the circulation generated by a sink is of opposite sign. The β-plume circulation is created by long Rossby waves emitted by the perturbation. The β-plume theory was further developed by Davey and Killworth[2] and Rhines[3] for the β-plane. A linear solution on the polar β-plane was presented in Ref. 4.

An interesting example of a β-plume in the zonal current was considered by Kaspi and Schneider[5]. Oceans provide spatially localized heating of the atmosphere eastward of the continents of North America and Asia due to the transport of warm water by the western boundary currents Gulf Stream and Kuroshio. The atmospheric circulation at the latitudes where this heating occurs contains a jet stream, the eastward zonal flow. It turns out that the Rossby waves emitted by the perturbation due to the heating are faster than the jet stream and are able to propagate upstream, in the westward direction. They form a β-plume which, surprisingly, causes the air over the eastern boundaries of the continents to be colder than the air above the ocean at similar latitudes.

Convective storms on Saturn, (also called Great White Spots) observed with regular intervals of approximately 30 years, (one Saturnian year) provide an example of the buoyancy sources embedded within zonal currents on a grander scale. The most recent storm was observed in 2010 by ground-based telescopes and the Saturn –orbiting Cassini spacecraft[6-8]. The storm was first detected as a small white spot (visible due to ammonia ice clouds) which grew rapidly in size and shed a tail of white clouds in the eastern direction. Observations and numerical modelling[7] show that the storm occurred due to upwelling of buoyant gases from a source located deep in the atmosphere. Zonal circulation at the latitude of the storm included two strong eastward jets on the northern and southern flank of the storm. The center of the storm was located within a relatively weak westward jet. It was hypothesized, however, that the source of the storm was embedded in a stronger westward flow in deeper layers such that the head of the storm appeared to drift westward relative to the top level westward jet. In contrast to the above example involving atmospheric flow on Earth, no westward propagation of a β-plume was observed on Saturn.

In what follows, we describe laboratory experiments on a rotating table with a setup modelling a polar β-plane. Convective flow is generated by an electric heater while background zonal jets are generated by a sink. The purpose of these experiments was to investigate the properties of the flow and to locate the boundary between regimes in the control parameters space. The two extreme regimes of the flow are the case where a (westward-propagating) β-plume forms and the case where the eastward transport due to the background eastward jet dominates. In our work we build upon work by Afanasyev et al.[4] and by Slavin and Afanasyev[9]; the latter found evidence for the formation of β-plumes in a setup which included an extended (along the radial direction in a circular tank) heater without any pre-existing background flow. The flows generated by the heater on the polar β-plane were investigated using laboratory experiments and numerical simulations. Zonal currents generated by sinks/sources were discussed in a pioneering paper by Armi[10] where the author developed a hydraulic theory to describe the nonlinear zonal currents and tested his theoretical results against laboratory measurements. In the present paper, we also perform a consistency check with the hydraulic theory.

In Sec. II of this paper we describe the setup of our apparatus as well as the optical altimetry technique used to measure the gradient of the surface elevation field, from which we obtain the velocity and vorticity fields. In Sec. III the results of the experiments and their analyses are reported. Concluding remarks are given in Sec. IV.

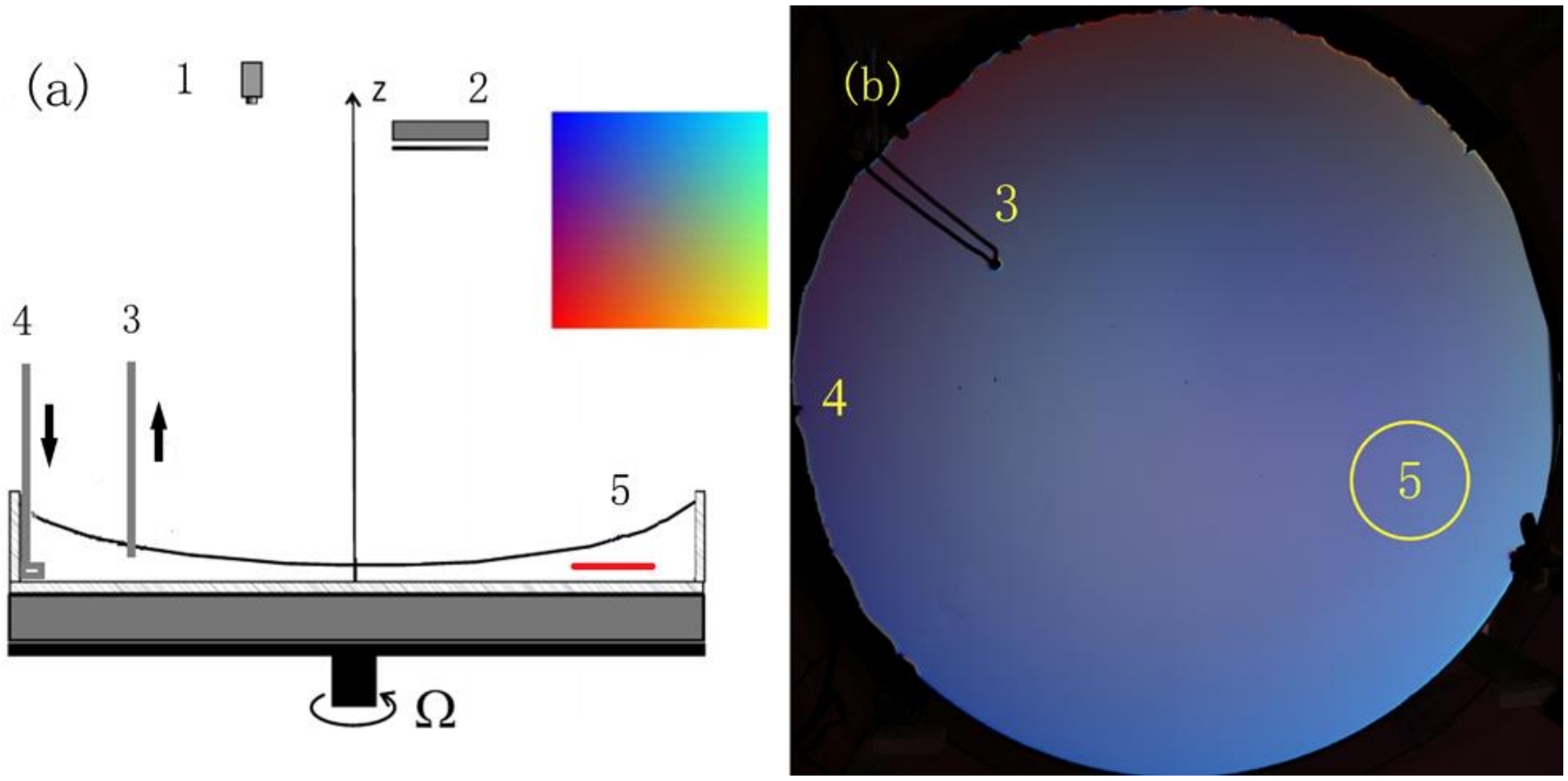

Figure 1. (Color online) Sketch of the experimental setup (a) and a snapshot of the tank rotating at “null” speed (b): video camera (1), high brightness display showing the color mask (2), fluid is pumped out of the tank through a thin tube (3) and returned through a tube (4), electric heater (5).

## II. LABORATORY APPARATUS AND TECHNIQUE

The experiments were performed in a cylindrical tank of radius $R$ = 55 cm installed on a rotating table (Fig. 1). The tank was rotated in an anticlockwise direction at a rate of $\Omega = 2.32$ rad s$^{-1}$. The depth $h$ of the rotating fluid with free surface varies quadratically with the distance $r$ from the axis of rotation

$$h(r) = H_0 + \frac{\Omega^2}{2g}\left(r^2 - \frac{R^2}{2}\right), \tag{1}$$

where $H_0$ = 10 cm is the depth of the layer in the absence of rotation and $g$ is the gravitational acceleration. The dynamical equivalence of the varying depth of the layer to the quadratically varying Coriolis parameter, $f = f_0 - \gamma r^2$ in the polar β-plane approximation results from the conservation of the potential vorticity (PV) defined as $q = (\zeta + f)/h$. Here $f_0 = 2\Omega$, $\zeta$ is the vertical component of the relative vorticity and $h$ is given by (1). Assuming that the percentage change in $h$ is small and that $\zeta$ is much smaller than the background vorticity $f_0$ (i.e. the Rossby number, Ro = $|\zeta| / f_0 \ll 1$) we obtain the following approximate expression for PV

$$q \approx \frac{1}{H_0}\left[\zeta + 2\Omega\left(1 - \frac{\Omega^2}{2gH_0}\left(r^2 - \frac{R^2}{2}\right)\right)\right]. \qquad (2)$$

We define the laboratory polar β-plane with parameter $\gamma$ given by

$$\gamma = \Omega^3 / (gH_0). \qquad (3)$$

In our experiments $\gamma$ = $1.3\times10^{-3}$ $s^{-1}cm^{-2}$. For the purpose of comparison with the previous results obtained in a β-plane approximation we can also introduce a local β-plane. An approximate expression for the depth of the fluid in the vicinity of radius $r_0$ is given by

$h(r) \approx h(r_0) - (\Omega^2 r_0 / g)y$, such that the PV can be written as[11]

$$q \approx \frac{1}{h(r_0)}\left[\zeta + 2\Omega\left(1 + \frac{\Omega^2 r_0}{gh(r_0)}y\right)\right]. \qquad (4)$$

Here the y-axis of the local Cartesian system with the origin at $r_0$ is directed towards the pole. The equivalence with the linearly varying Coriolis parameter, $f = f_0 + \beta y$, can be established by defining the β-parameter as

$$\beta = 2\Omega^3 r_0 / gh(r_0)\,. \tag{5}$$

In our experiments the value of $\beta$ at the radius of the heater is $\beta = 0.1\ \mathrm{s^{-1}cm^{-1}}$.

Flows in the tank were induced by an electric heater and a source-sink arrangement. Their positions in the tank are shown in Fig. 1. The heater was used to create an upwelling of warm water from the bottom to the surface. A heater of radius $r_h$ = 6.5 cm was placed 1 cm above the bottom of the tank at $r$ = 39 cm (midlatitudes of the tank). The maximum power of the heater was $Q$ = 960 W. The power of the heater was varied between the experiments in the range between 10% and 100% of the maximum value. Following Ref. 10 we used a sink to create an eastward (counterclockwise) zonal flow. Water was pumped out of the tank by an electric pump and then returned back to the tank through the source located near the wall. The source was equipped with a diffuser in order to minimize its effect on the circulation in the tank. The intensity of the pump was varied between the experiments to obtain the desired velocity of the zonal flow.

The Altimetric Imaging Velocimetry (AIV) system was used to observe perturbations of the surface topography and to measure two components of the gradient $\nabla\eta$ = $(\partial\eta/\partial x,\ \partial\eta/\partial y)$ of the surface elevation $\eta$ in the horizontal plane ($x$, $y$). Here we describe briefly the AIV technique (for more details see Ref. 11). The system includes a 5 Mpix video camera capable of recording at a rate of up to 10 frames per second and a high brightness computer monitor acting as a light source (Fig.1). A color mask shown in inset in Fig. 1a was displayed on the monitor. The video camera observes the reflection of the color mask in the surface of the water. At a certain rotation rate of the tank the focal distance of the paraboloid mirror, formed by the rotating water is such that the camera sees the surface uniformly illuminated by nearly one color (that of the center of the mask). We call this rotation rate a null rotation rate and all altimetric measurements are performed at this rate. A snapshot of the surface at null rotation rate is shown in Fig. 1 b. Note that the field in Fig. 1 b is not completely uniform across the tank and some variation of color remains. This second-order effect is discussed in Ref. 11 and is accounted for in the AIV software. The value of the null rotation rate is determined by the geometry of the AIV setup. When the surface of water is perturbed by the flow the slopes of the surface elevation field change the angles of reflection. As a result, different areas of the light source are reflected. This creates multicolor images of the surface which are used then to visualize the flow and to measure the components of $\nabla\eta$.

The surface (barotropic) velocity of the flow can be determined from the measured gradient of surface elevation using shallow water and quasigeostrophic approximations which yield

$$\mathbf{U} = \frac{g}{f_0}(\mathbf{n}\times\nabla\eta) - \frac{g}{f_0^2}\frac{\partial}{\partial t}\nabla\eta - \frac{g^2}{f_0^3}J(\eta,\nabla\eta), \tag{6}$$

where **U** is the horizontal velocity vector, **n** is the vertical unit vector and $J(A,B) = \frac{\partial A}{\partial x}\frac{\partial B}{\partial y} - \frac{\partial B}{\partial x}\frac{\partial A}{\partial y}$ is the Jacobian operator. Two altimetry images of the flow separated by a short time interval are required to calculate the velocity using Eq. 6. The first term on the RHS of Eq. 6 can be easily recognized as the geostrophic velocity. The second and third terms are due to transient and nonlinear effects. The relative importance of the latter terms is determined by the temporal Rossby number $Ro_T = 1/(f_0 T)$ and the Rossby number $Ro = U/(f_0 L)$ respectively. Here $T$ is the time scale of the flow evolution, while $U$ and $L$ are velocity and length scales of the flow. In the experiments we measure a “true” field of the gradient of the surface elevation (in fact the pressure gradient) but the calculated velocity field is more accurate when the dimensionless parameters $Ro_T$ and Ro are small.

In some experiments an infrared (IR) camera (Flir E30) with a 120 x 160 pixel array was used to visualize and measure the temperature field at the surface of water in the vicinity of the heater. The velocity due to baroclinic effects can be determined from the measured temperature field. The thermal wind equation allows us to relate the vertical shear of horizontal velocity with the horizontal gradient of density at the surface,

$$\frac{\partial \mathbf{U}}{\partial z} = -\frac{g}{\rho_0 f_0}\left(\mathbf{n}\times\nabla\rho\right). \qquad (7)$$

In a first approximation, the density $\rho$ is linearly proportional to the perturbation temperature $T$, $\rho = \rho_0(1 - \alpha T)$. Here $\alpha$ is the thermal expansion coefficient and $\rho_0$ is the unperturbed density. Eq. (7) then takes the form

$$\frac{\partial \mathbf{U}}{\partial z} = \frac{g\alpha}{f_0}\left(\mathbf{n}\times\nabla T\right). \qquad (8)$$

Table I. Parameters of laboratory experiments.

| Expt. ID | Heater intensity (% of max.) | $U_{east}$ (cm/s) | | | | | | | |
|---|---|---|---|---|---|---|---|---|---|
| | | A | B | C | D | E | F | G | H |
| 1 | 100 | 0.00 | 0.15 | 0.30 | 0.38 | 0.54 | 0.36 | 0.64 | 0.69 |
| 2 | 80 | 0.00 | 0.17 | 0.27 | 0.39 | 0.55 | 0.37 | 0.48 | 0.76 |
| 3 | 65 | 0.00 | 0.21 | 0.23 | 0.43 | 0.55 | 0.37 | 0.44 | 0.75 |
| 4 | 45 | 0.00 | 0.24 | 0.23 | 0.45 | 0.61 | 0.36 | 0.47 | 0.78 |
| 5 | 25 | 0.00 | 0.23 | 0.29 | 0.45 | 0.61 | 0.52 | 0.47 | 0.77 |
| 6 | 10 | 0.00 | 0.22 | 0.29 | 0.44 | 0.56 | 0.50 | 0.50 | 0.78 |

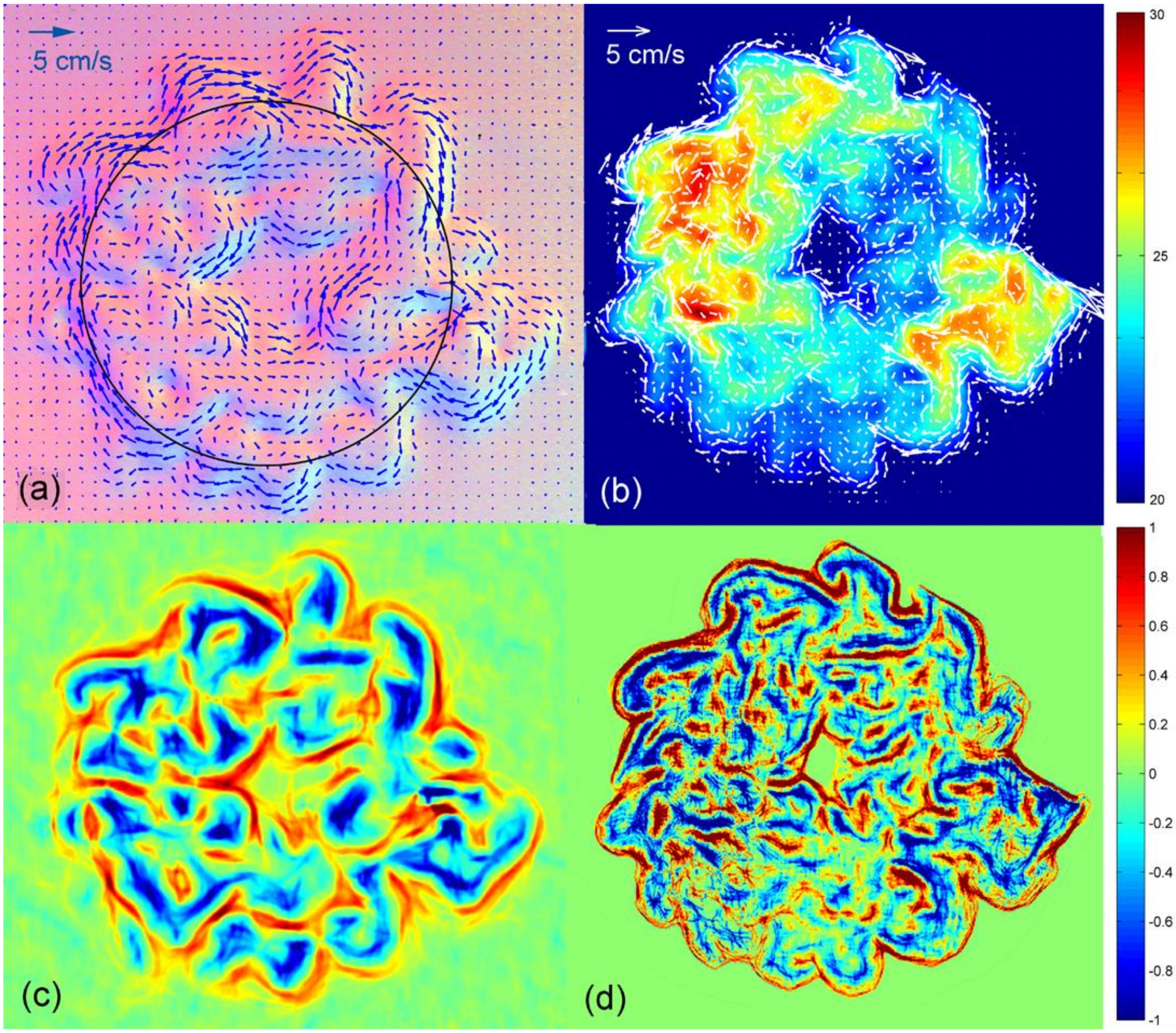

Fig. 2 (Color online) Flow above the heater at $t$ = 28 s after the heater was turned on in experiment A1 (Table 1). (a) Barotropic velocity vectors superposed on the altimetric image of the flow. (b) Baroclinic velocity vectors superposed on the temperature field measured by IR camera, color scale shows temperature in $^0$C. (c, d) Relative vorticity fields calculated using the barotropic or baroclinic velocity respectively. The color scale shows dimensionless vorticity $\zeta/f_0$ in the range from –1 (blue, anticyclonic) to 1 (red, cyclonic) . Black circle in panel (a) shows the outline of the heater.

## III. EXPERIMENTAL RESULTS AND ANALYSES

Herein we describe experiments with varying heater power and with varying strength of the zonal flow induced by the sink. Table 1 shows the main control parameters of the experiments namely the intensity of the heater (in % of maximum) and the velocity, $U_{east}$ of the eastward background jet (the details of the background flow created by the sink will be discussed later). In what follows particular experiments from Table 1 will be referred to as e.g. “A1” or “D6”.

The first series of experiments (A1-6 in Table 1) was performed without any background zonal flow in order to study the flow generated by the heater only. This series provided a reference point for further experiments with background zonal flow. Observations show that warm water from the heater rises to the surface and creates a turbulent flow with multiple eddies. Figure 2 shows a magnified picture of the surface area above the heater shortly after the start of the experiment. The surface velocity field measured by AIV (Fig. 2 a) revealed an intense jet-like flow at the periphery of the warm lens created by the spreading warm water. This is a typical frontal jet which flows anticyclonically (clockwise). Another frontal jet which flows cyclonically (counter clockwise) can be observed around the central part of the area where there is no warm water initially. The periphery of the lens is subject to frontal/baroclinic instability[12,13] and multiple meanders and eddies form there. Multiple eddies are also created inside the lens. The vorticity map in Fig. 2 c shows both cyclones (red) and anticyclones (blue). It is interesting to note that the anticyclones are mainly in the form of elongated bands while cyclones are often of circular form. Circular cyclones are most likely associated with the areas of local downwelling of relatively cool water from the surface where convergence of horizontal velocity results in cyclonic circulation. Figure 2 b shows a surface temperature map measured by the IR camera together with the baroclinic velocity field. The vertical shear of baroclinic velocity was calculated using Eq. (8). The shear was then multiplied by $h/2$, the half-depth of the fluid, to obtain an estimate of velocity. The value of the effective thickness of the warm lens that was chosen for this estimate will be discussed later. A comparison of Fig. 2 a and Fig. 2 b shows that the direction and magnitude of the barotropic (AIV) and baroclinic (IR) velocities are generally similar. The differences between these two fields can be revealed by comparing the corresponding vorticity fields in Fig. 2 c and d. The baroclinic vorticity field consists mainly of narrow bands of paired positive and negative vorticity which correspond to intense frontal jets while barotropic vorticity is much less concentrated.

Flows created by a buoyancy source on the f-plane were studied in detail in laboratory experiments by Fernando et al.[14] and Maxworthy and Narimusa[15]. The results of these experiments allow us to establish a consistent physical picture of the initial development of our flow. Refs. 14 and 15 showed that a mixed layer forms above the buoyancy source. The flow in the mixed layer is turbulent and is not affected by the background rotation of the fluid. Beyond the mixed layer the rotation becomes important and columnar vortices are formed there. A transition depth $z_c$ between these regions is determined by the values of the buoyancy flux $B_0$ and the Coriolis parameter $f_0$ such that $z_c \approx 13\left(B_0^{1/2} / f_0^{3/2}\right)$. The buoyancy flux is defined as $B_0 = \alpha g Q/(C_p A_0)$, where $Q$ is the power of the heater, $C_p$ is the fluid's volumetric heat capacity at constant pressure and $A_0 = \pi r_h^2$ is the area of the heater. In our experiments, the buoyancy flux varied in the range between 0.1 and 0.41 $cm^2/s^3$ and the transition depth was quite small, $z_c$ = 0.4 – 0.8 cm. Thus the most of the layer above the heater is occupied by rotationally dominated convection. The results described in Refs. 14 and 15 give the upward speed of propagation of warm water in this region in the form

$$w \approx 1.0\left(B_0 / f_0\right)^{1/2} . \tag{9}$$

Warm water rising to the surface spreads in the horizontal direction forming a layer of thickness $h_1$ (note that $h_1$ is not the change in elevation of the surface but the thickness of the warm upper layer). Equating the volume flux from the heater with the rate of change of the volume of the upper layer we obtain

$$wA_0 = h_1 \frac{dA}{dt} , \tag{10}$$

where $A$ is the area of the layer. We measured the radius of the expanding lens of warm water and calculated its area $A$ (Fig. 3 a). The measurements show that the area of the lens increases at an approximately constant rate initially before stabilizing at some constant value. The measured rate of change of the area varies linearly with $w$ (Fig. 3 b) such that $dA/dt \approx 34w$. Substitution into (10) gives $h_1 \approx 0.45h$. Thus the thickness of the expanding upper layer is approximately one half of the total depth $h$. This result is not surprising in the context of exchange gravity flows[16,17].

After a relatively short initial period when the warm water lens expands in an almost axisymmetric manner as if on the f-plane, the influence of the β-effect on the flow evolution becomes evident. Figure 4 shows the development of the flow at two different times, at the beginning and at the end of the experiment. Snapshots of the flow visualized by AIV are shown in Figs 4 a and b together with the distributions of barotropic azimuthal velocity (see insets). The surface elevation field $\eta$ obtained by integration of the measured gradient in x- and y-directions is given in Figs 4 c and d. The fields in Fig. 4 show that zonally-elongated barotropic Rossby waves are emitted westward (clockwise) from the heater. The observations of zonally elongated waves are consistent with the previous experimental and numerical results of Afanasyev et al.[4] and Slavin and Afanasyev[9] where different buoyancy sources were used. In Ref. 9, it was shown that the zonal phase speed of the waves was in good agreement with that predicted by the dispersion relation for the barotropic Rossby waves in a cylindrical container. The Rossby waves are emitted at the front of the expanding baroclinic layer. Consistent with the previous observations[4, 9] the waves are emitted by the individual baroclinic meanders rather than by the entire region above the heater as a single source. In Fig. 4 c, for example, we can clearly see the elongated region of relatively high surface elevation protruding from the most prominent meander/eddy at the western edge of the heater. The ridge at the surface is followed by the valley to the south of the meander and then by another ridge extended from a meander at the southern edge of the heater. Note that if the size of the source is comparable with the baroclinic radius of deformation or smaller, a single β-plume is formed. It was demonstrated in idealized experiments where a source was in the form of a thin tube with a diffuser (see Figs 4 and 5 in Ref. 4). The emission of zonally elongated Rossby waves results in the formation of almost zonal currents/jets of alternating eastward and westward direction (Fig. 5). It is interesting to note also an alternating pattern of highs and lows of surface elevation to the west of the heater (between approximately 12 and 3 o'clock in Fig. 4 d). This pattern indicates a Rossby wave of relatively short zonal wavelength which is capable of propagating to the west of the perturbed area.

Long Rossby waves of almost zero-frequency establish a stationary circulation in the form of β-plumes. Each β-plume emitted by a source is a ridge of surface elevation extended to the west of the source. The ridge corresponds to a gyre consisting of two jets flowing in opposite directions. The jets are connected within the area of the source. The distributions of zonal velocity in Fig. 4 indicate that two β-plumes are created in this particular experiment. A linear theory of β-plumes generated by sources of heat was given in Refs. 1-3. Ref. 3 considered a plume generated in a single layer by localized buoyancy forcing/upwelling on a β-plane. A similar approach was adopted in Ref. 4 to describe a β-plume on the polar β-plane. Since in these experiments we employ the same circular polar β-plane geometry we can use the solution given in Ref. 4. The steady state solution predicts that the elevation of the surface $\eta$ created by the perturbation is uniformly extended westward from the perturbation area such that

$$\eta(r,\theta) = \frac{1}{2R_d^2\gamma}\int_{\theta}^{\theta_E} w(r,\theta')d\theta' = \frac{4}{f_0}\int_{\theta}^{\theta_E} w(r,\theta')d\theta' \, , \tag{11}$$

where $(r, \theta)$ is the polar coordinate system with the origin at the center of the tank, $R_d = (gh)^{1/2}/f_0$ is the barotropic radius of deformation and the upper integration limit $\theta_E$ is the eastern boundary of the perturbation area. Here we also used Eq. (3) for $\gamma$. Thus, the surface elevation is in the form of a "ridge" stretching along the zonal direction. Geostrophy predicts that two jets will flow in opposite directions on the northern and southern slopes of the ridge with velocity given by

$$\mathbf{U} = \frac{g}{f_0}\left(\mathbf{n}\times\nabla\eta\right) = \frac{g}{f_0}\left(-\frac{1}{r}\frac{\partial\eta}{\partial\theta}, \frac{\partial\eta}{\partial r}\right). \tag{12}$$

The radial component of velocity exists only within the area of the perturbation (non-zero $w$) and serves to connect the zonal jets. The radial transport is therefore the same as the zonal transport in each of the jets. The radial velocity is given by the following simple expression

$$U_r = -\frac{4g}{rf_0^2} w . \qquad (13)$$

Similarly, the azimuthal velocity is given by

$$U_\theta = \frac{4g}{f_0^2} \frac{\partial}{\partial r} \int_{\theta}^{\theta_E} w(r, \theta') d\theta' . \qquad (14)$$

Thus, both components of velocity in the β-plume are proportional to the vertical velocity $w$ in the perturbation area. We measured the profiles of the azimuthal velocity along the radial line going through the western edge of the heater (solid line in Fig. 4) when β-plumes were well formed. The instantaneous profiles were then averaged over a significant period of time (400 s) to obtain a mean profile of the azimuthal velocity (Fig. 5 a). The peak values of the velocity in the westward jet located at the southern edge of the heater for experiments A1- 6 are plotted in Fig. 5 b against the vertical velocity $w$ calculated from the buoyancy flux $B_0$ in each experiment. A linear fit gives $U_{\text{west}} = 2.1w$, where $U_{\text{west}} = -U_\theta$. Note that the factor $4g/(rf_0^2)$ in (13) is approximately 4.7, which is not unreasonable compared to the measured value of 2.1. Of course, these values can be compared only within a factor of order of unity because we do not know the exact distribution of the vertical velocity $w$ across the area above the heater.

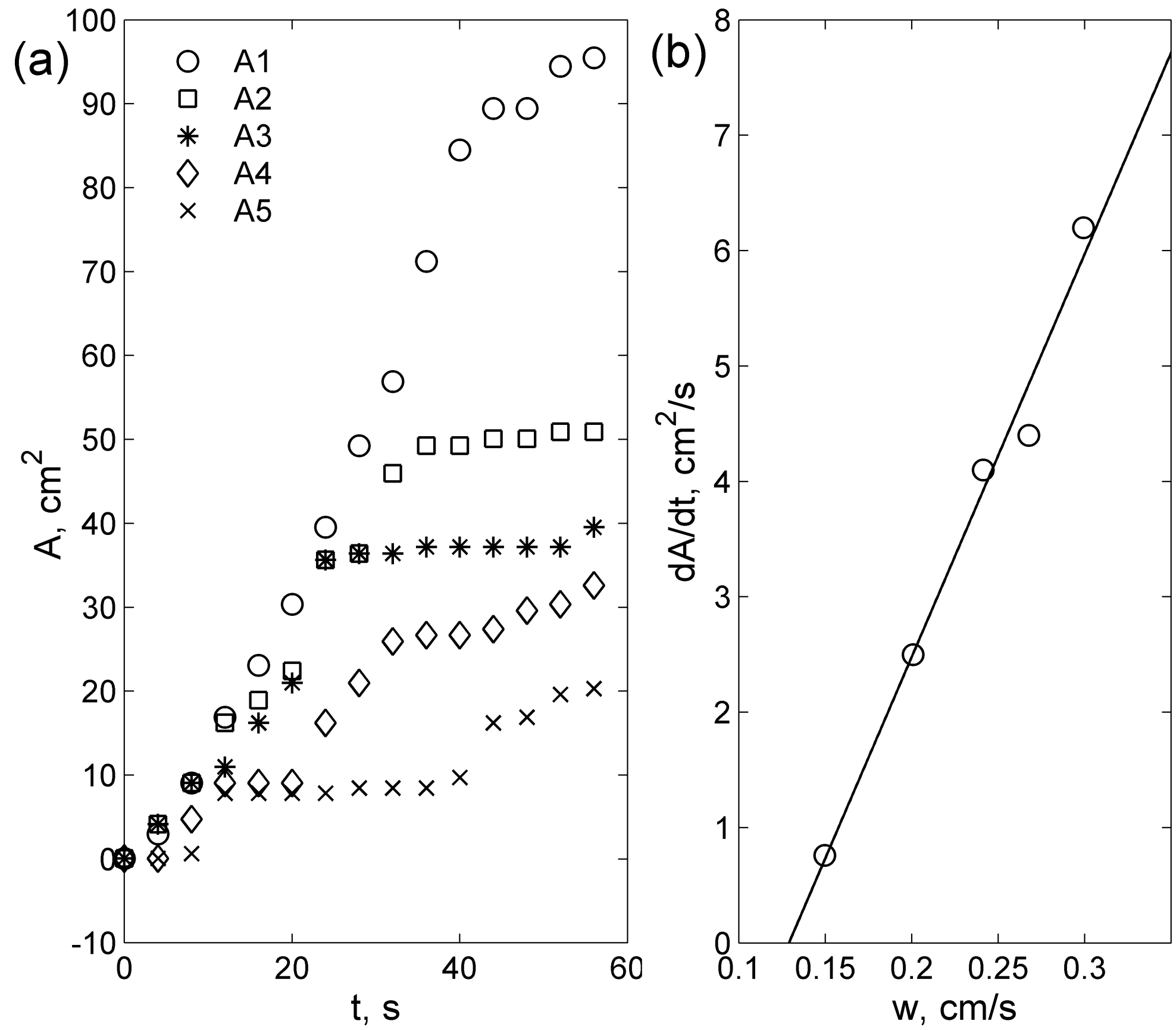

Fig. 3. Expansion of the area of the warm water lens above the heater in experiments A1-5 (Table 1). (a) *A* as a function of time. (b) Rate of change of the area in the initial period of flow evolution against the upwelling velocity *w*. The solid line shows a linear fit.

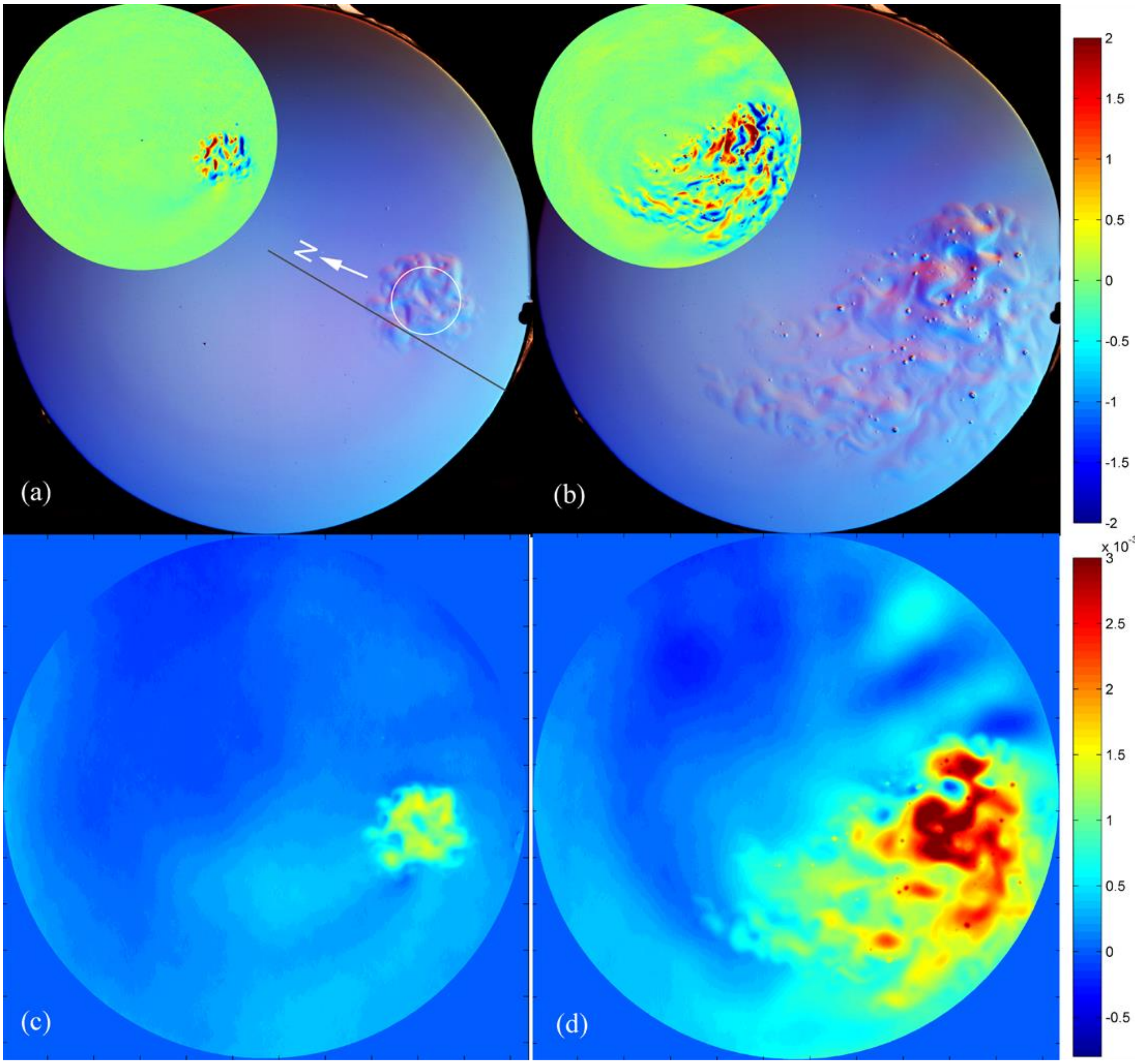

Fig. 4 (Color online) Development of a β-plume in experiment A1(Table 1): (a, b) altimetric images and (c, d) surface elevation varying in the range from $-8\times10^{-4}$ cm (blue) to $3\times10^{-3}$ cm (red) (see color scale on the right). The fields are observed at $t$ = 53 s (a, c) and $t$ = 433 s (b, d). The insets in (a) and (b) show the azimuthal velocity field varying in the range from –2 cm/s (blue) to 2 cm/s (red) (see color scale on the right). White circle indicates the outline of the heater. The solid line drawn in the radial direction through the western edge of the heated area indicates where the profiles of azimuthal velocity are measured (see next figure). White arrow shows the local direction to the North (towards the center of the tank).

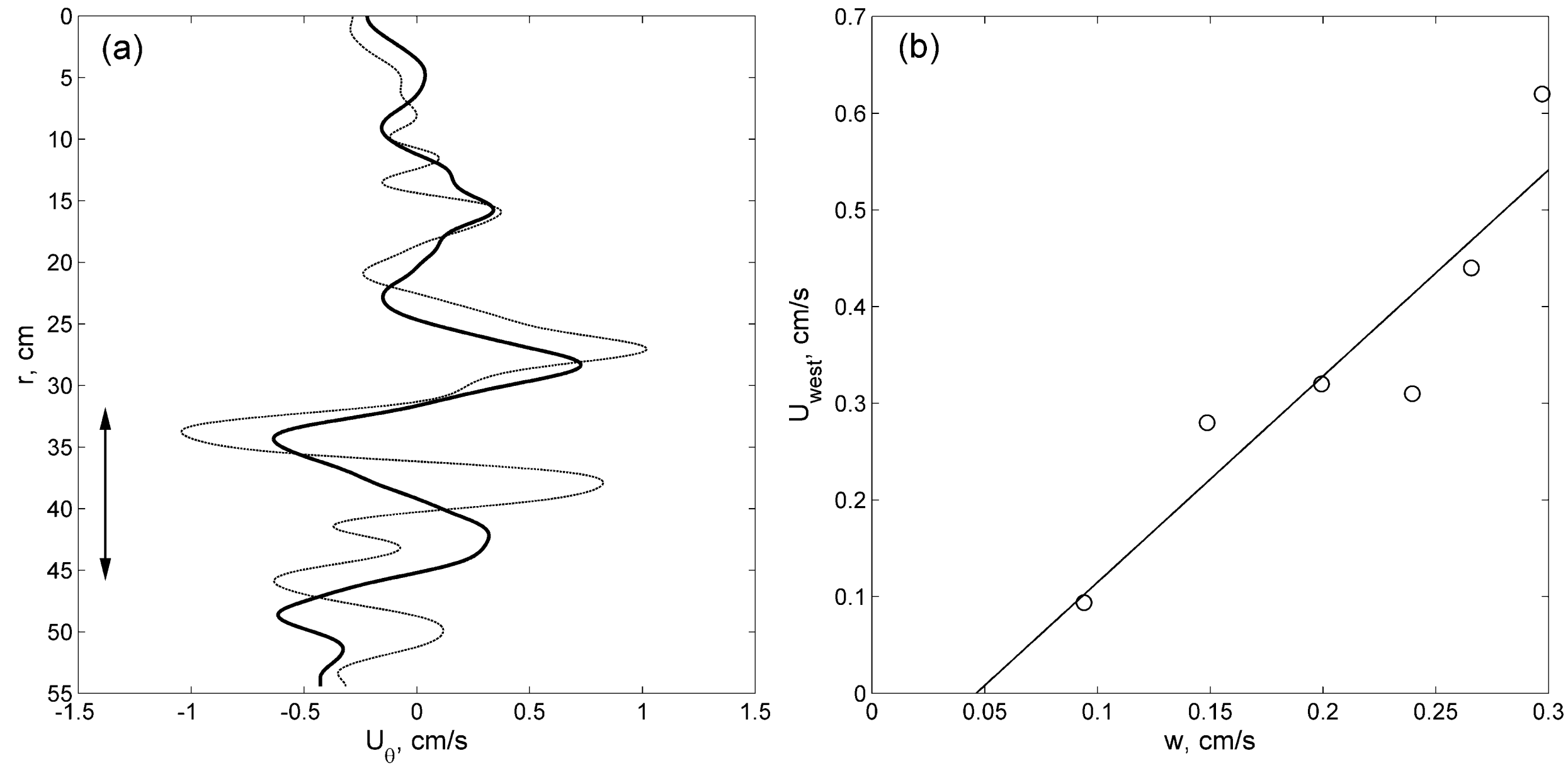

Fig. 5. Zonal currents due to β-plume circulation generated by the upwelling above the heater in experiment A1(Table 1). (a) Profiles of the azimuthal velocity along the radial line at the western edge of the heater (as shown in Fig. 4). Positive (negative) velocity corresponds to eastward (westward) flow. The solid black line shows the time mean profile while the gray line shows an instantaneous profile corresponding to time displayed in Fig. 4 b. The double arrow indicates the radial extent of the heater. (b) Peak westward velocity against the upwelling velocity *w*. The solid line shows a linear fit.

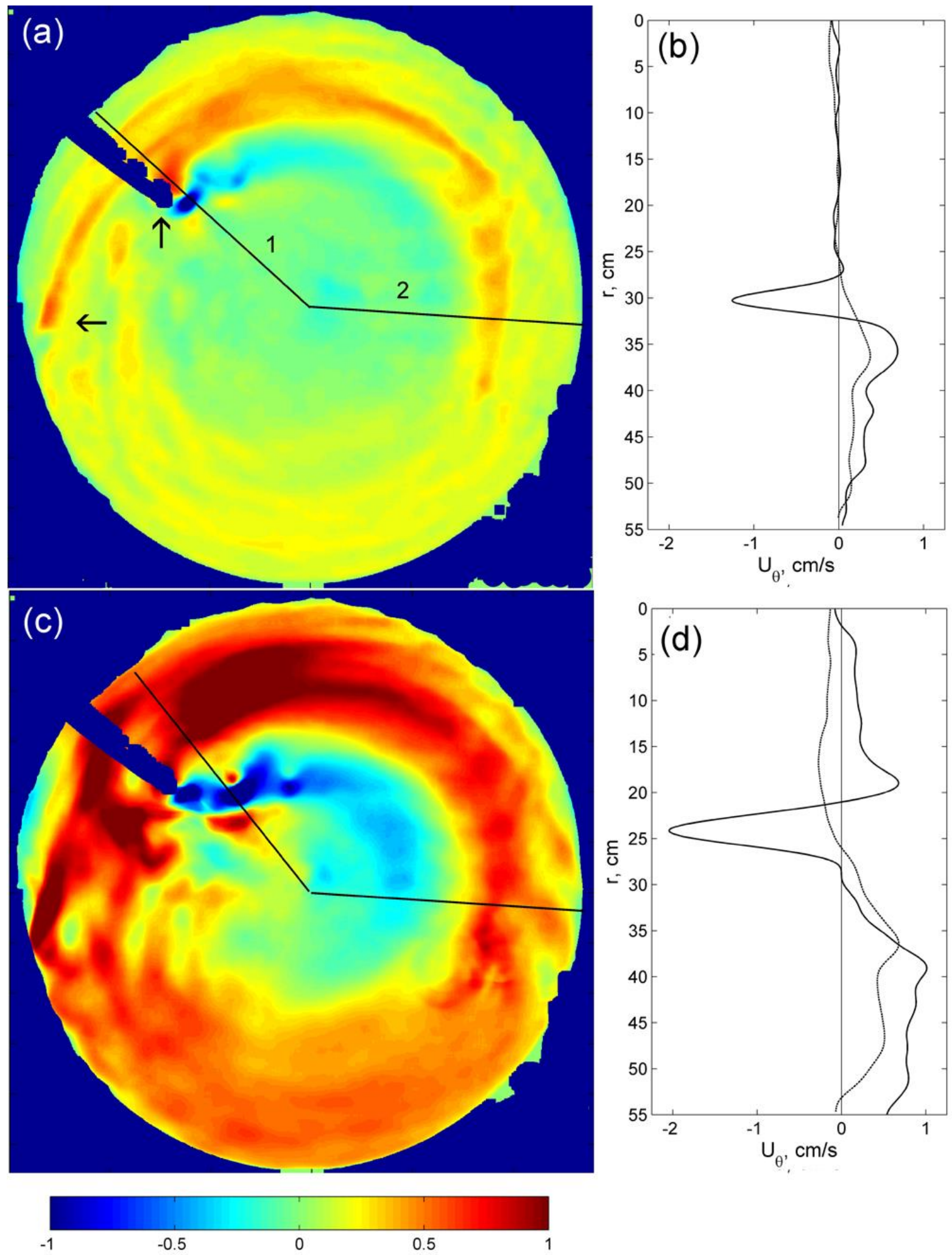

Fig.6 (Color online) Azimuthal velocity of the flows created by a sink in experiments B1 (a, b) and H1 (c, d) before the heater was turned on. The color scale shows azimuthal velocity in cm/s. Positive (negative) velocity corresponds to eastward (westward) flow. Profiles in (b) and (d) are measured along the radial lines 1 and 2; solid black lines show velocity along line 1 in the vicinity of the source where $Ro_\beta \approx 1$ while the grey lines show velocity along line 2 at the eastward edge of the heater. The arrows in (a) indicate the positions of the sink (at approximately 10 o'clock) and the diffused source (at approximately 9 o'clock).

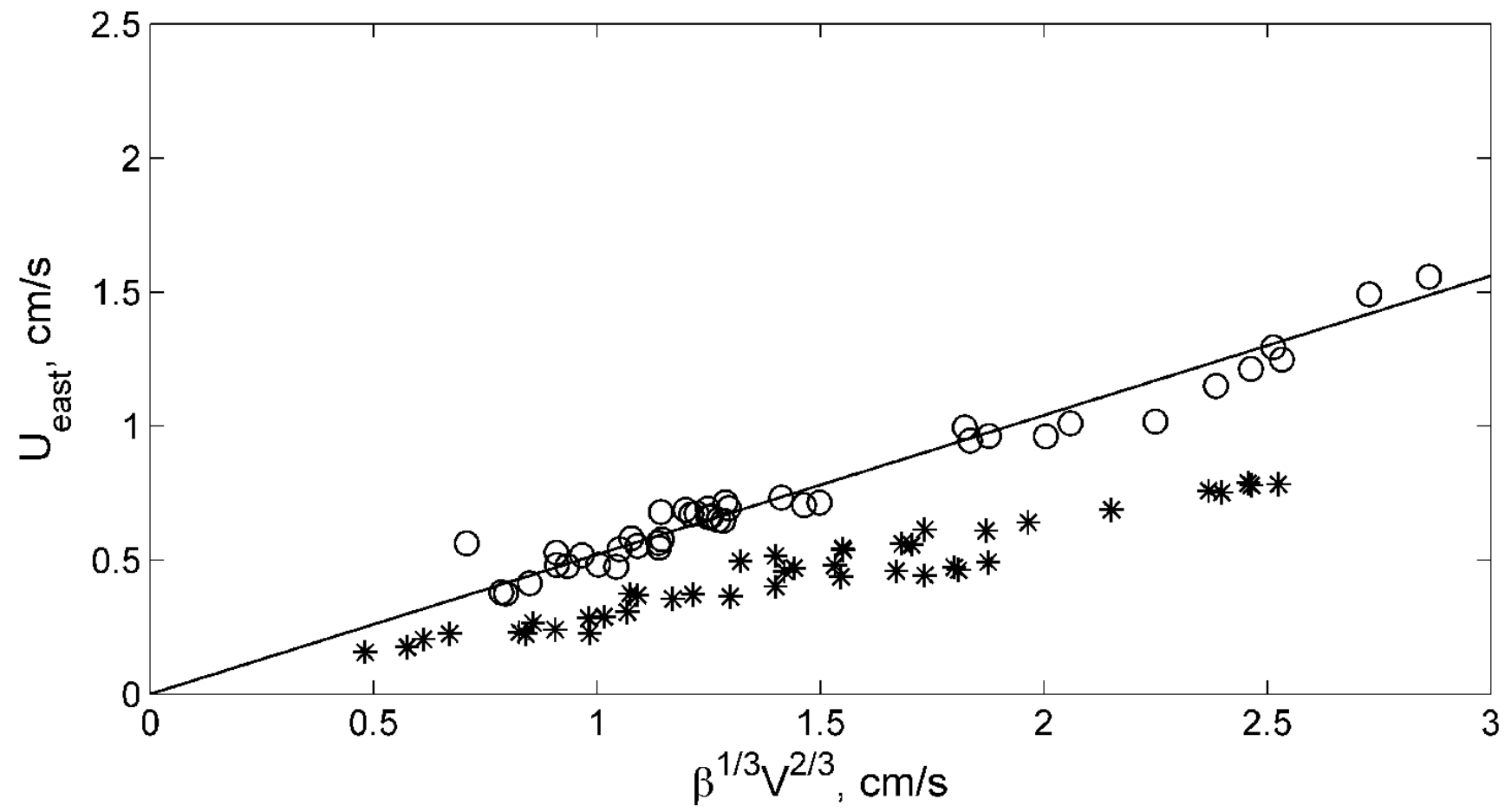

Fig. 7. The peak velocity of the eastward zonal current created by the sink as a function of $\beta^{1/3}V^{2/3}$. Circles show the velocity measured in the vicinity of the sink where $Ro_\beta \approx 1$ (see Fig. 6) while asterisks show the velocity measured near the eastward edge of the heater. The solid straight line has slope 0.52 as in the theoretical expression (15).

Let us now discuss the flows generated by a sink on the polar β-plane. In experiments B-H (Table 1) the sink was started first and then enough time was given for an almost stationary circulation to form before the heater was started. Figure 6 shows two examples of the flow generated by the sink working at a relatively low rate (Fig. 6 a, b) and a high rate (Fig. 6 c, d). The distributions of the azimuthal velocity, as well as the profiles measured along the line drawn radially through the eastward boundary of the heater (solid line in Fig. 6 a) show two distinct zonal bands. The band of eastward flow starts at the latitude of the sink and extends southward almost to the boundary of the tank, while the band of westward flow extends northward from the sink. Note that the radial position of the sink was chosen such that the heater is located right in the eastward band. The sink effectively generates a β-plume, but with circulation opposite to that generated by the source i.e. eastward south of the sink, and westward north of it. Note that the actual picture of the circulation in the tank is however somewhat different from that of a single β-plume due to the sink. It is complicated by the presence of the diffused source at the wall (at approximately 9 o'clock in Fig. 6). The source generates its own plume but the eastward flow of this plume connects with that of the sink while the westward flow is effectively suppressed by the wall of the tank (Fig. 6). Some transient effects in the form of small (barotropic) cyclonic vortices separating from the sink can also be observed but these transients decay and become negligible before they reach the longitude of the heater.

β-plumes at low Rossby numbers can be described by a linear theory[1-4]. An alternative approach was offered by Armi[10] who used the analogy between the zonal currents on the β-plane and open channel hydraulics. Hydraulic theory shows that the eastward flow generated by a sink is supercritical in the vicinity of the sink and subcritical farther upstream (westward). The main control parameter is the Froude/Rossby number defined as $Ro_\beta = U_{east}/(\alpha'\beta a^2)$, where $U_{east}$ is the maximum velocity of the (eastward) zonal current, the coefficient $\alpha'$ is determined by the shape of the velocity profile ($\alpha'$ = 0.25 for a parabolic profile), and $a$ is the half-width of the current. In hydraulic flows terminology, a control section is the location where $Ro_\beta = 1$. To the west of the control section the flow is subcritical ($Ro_\beta < 1$) such that the Rossby waves are fast enough to propagate upstream and carry information from the control section. The theory given in Ref. 10 gives the relation between the peak velocity $U_{east}$ and the volume flux $V$ as

$$U_{east} = 0.52\beta^{1/3}V^{2/3}. \qquad (15)$$

The volume flux $V$ is defined as

$$V = \int_{-a}^{a} u(y)dy \qquad (16)$$

where $u(y)$ is the velocity profile in the zonal current in the local Cartesian coordinates $(x, y)$ with the origin on the axis of the jet. Experimental results obtained by Armi[10] showed reasonable agreement with the theoretical prediction (15). We performed similar measurements in all of our experiments (except A1- 6) to check the consistency of our results with the theory. The velocity $u(y)$ was measured along a radial line drawn sufficiently close to the sink such that $Ro_\beta \approx 1$ at this particular cross section (the location of the line was different in experiments with different strengths of the sink). These results are summarized in Fig. 7 where the peak velocity $U_{east}$ is plotted as a function of $\beta^{1/3}V^{2/3}$. The solid line shows the theoretical relation (15) and is in a good agreement with the experimental data. The data obtained along the line 2 (Fig. 6) at the eastward edge of the heater is also shown in Fig. 7. The values of the peak velocity $U_{east}$ and the volume transport $V$ at this cross section are lower than those at the control section. The values of the Froude/Rossby number are below unity, in the range between $Ro_\beta = 0.08$ and $Ro_\beta = 0.25$ in different experiments. The flow, as one might expect, is subcritical and approximately linear. Note that the volume transport in the eastward current upstream of the control section decreases with distance because of the recirculation occurring due to Ekman pumping/suction in the bottom boundary layer (see Fig. 1 in Ref. 1), and to lesser extent, due to the lateral exchange with the westward current in the north.

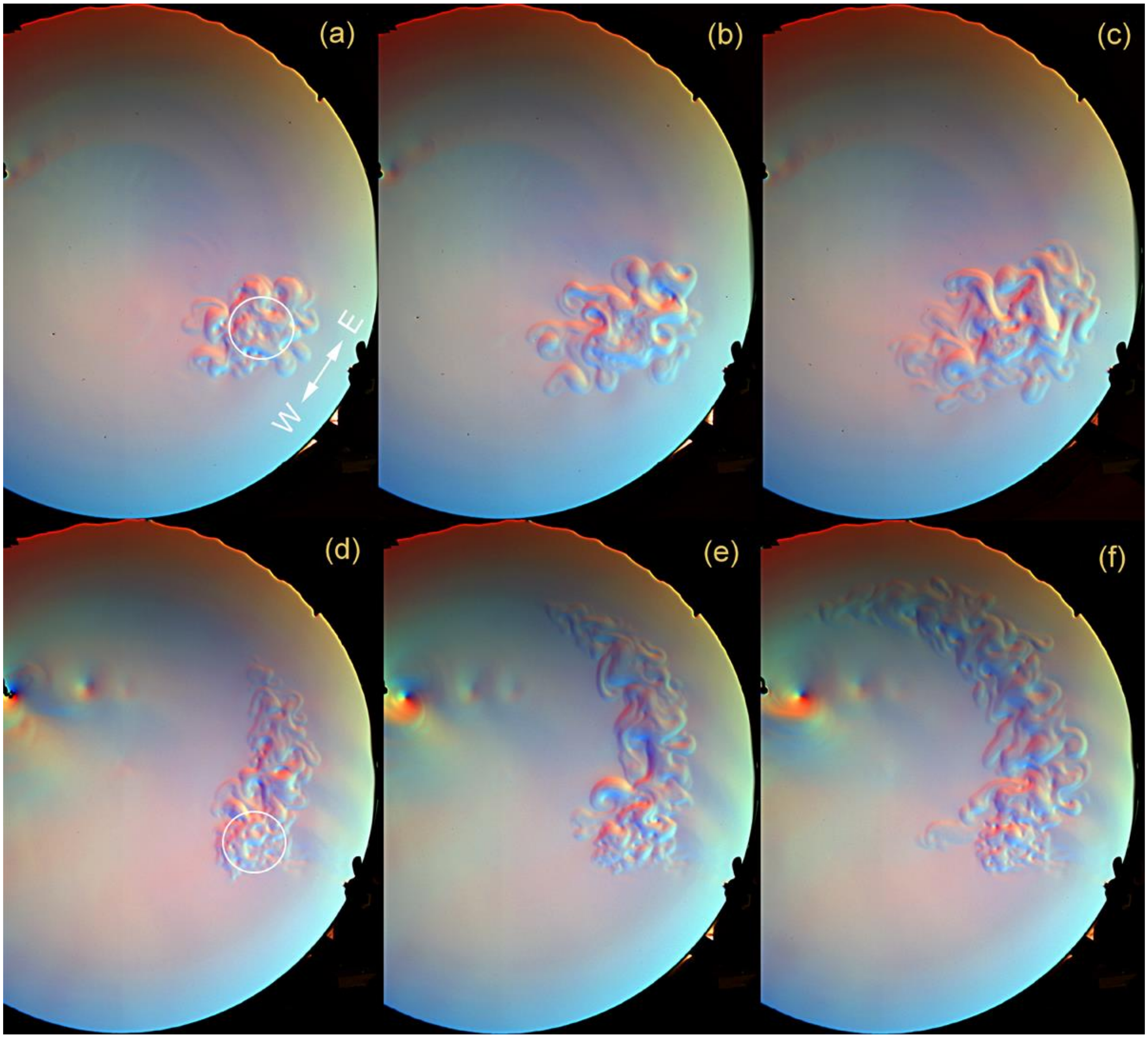

Fig. 8 (Color online) Altimetric images showing the evolution of the flow generated by a heater in the experiments B1 (a-c) and H1 (d-f) where the background zonal flow was weak or strong, respectively. The images correspond to times $t$ = 80 s (a, d), $t$ = 120 s (b, e) and $t$ = 160 s (c, f). White circles in panels (a) and (d) show the outline of the heater. The East-West direction is indicated by a white arrow in (a). $U_{east}$ = 0.15 cm/s and 0.69 cm/s in the experiments B1 and H1 respectively and $w$ = 0.3 cm/s in both experiments.

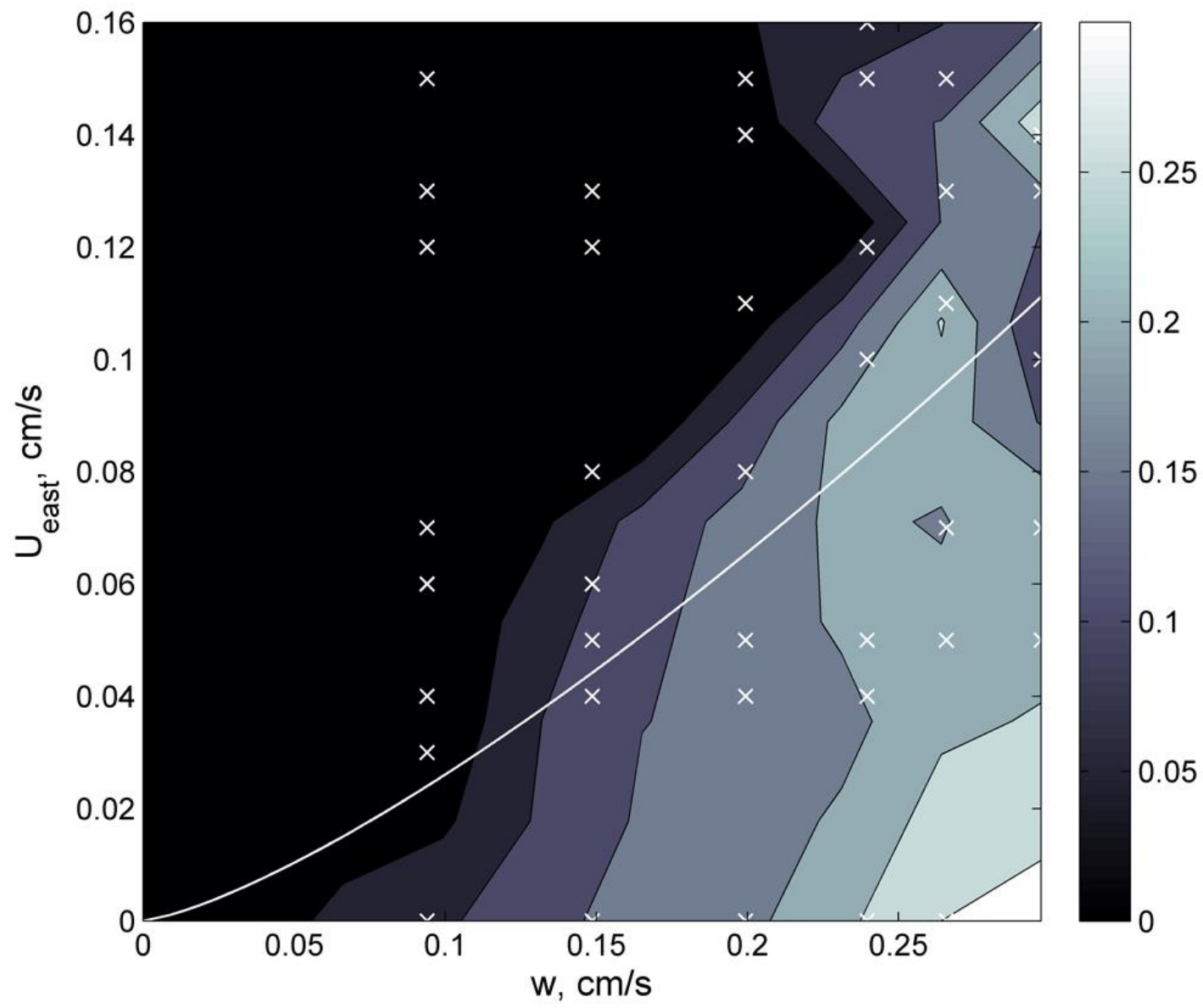

Fig. 9. Contour plot of $U_{\mathrm{plume}}$ for different values of $w$ and $U_{\mathrm{east}}$ . The gray scale shows the values of $U_{\mathrm{plume}}$ normalized by the speed of zonally elongated Rossby waves of meridional extent equal to the diameter of the heater, $c_{gx} = \beta(r_h/\pi)^2$. Crosses indicate the values of $U_{\mathrm{east}}$ and $w$ for each experiment where $U_{\mathrm{plume}}$ was measured. The solid white line shows Eq. (19).

After the zonal flow due to the sink was well-formed and stationary, the heater was switched on in experiments B-H (Table 1). Eastward transport due to the sink and westward β-plume flow due to the heater are competing effects in these experiments. Two sequences of AIV images of the flow in Fig. 8 show the two extremes in the flow regime. When the heater is relatively strong while the sink is weak, the flow is a β-plume (Fig. 8 a - c) similar to that in Fig. 4. When the ratio of the strength of the heater and the strength of the sink is reversed, the β-plume is not formed and warm water rising to the surface above the heater is carried to the east by the zonal flow (Fig. 8 d - f). In intermediate cases, both eastward and westward transport was observed. In order to find the boundary between the two regimes, we plot the distribution of the velocity $U_{\mathrm{plume}}$ of the westward propagation of the β-plume as a function of two control parameters, the vertical velocity $w$ (determined by the heater power) and the eastward jet velocity $U_{\mathrm{east}}$ (determined by the flux through the sink). Here $U_{\mathrm{east}}$ was measured at the eastern boundary of the heater at line 1 (Fig. 6 a) and $w$ was calculated using Eq. (9). The mean velocity of the westward propagation of the plume, $U_{\mathrm{plume}}$ was obtained by measuring the time it takes for the front of the plume to reach a longitude of $30^0$ west of the heater. Note that $U_{\mathrm{plume}}$ is positive for the westward propagation. Some uncertainty in these measurements was introduced by the fact that warm water from the heater gets entrained into the westward zonal current generated by the sink at latitudes to the north of the heater. It was sometimes difficult to separate the transport due to the β-plume from that due to the westward zonal current. This uncertainty was greater when the sink, and hence the zonal current, was stronger. The values of $U_{\mathrm{plume}}$ measured in each experiment for particular values of $w$ and $U_{\mathrm{east}}$ (indicated in Fig. 9 by crosses) were interpolated on a regular grid in the plane ($w$, $U_{\mathrm{east}}$) and showed as a contour plot. As one might expect, Fig. 9 shows that there is no westward propagation when the sink is relatively strong and the heater is relatively weak. Some insight into the dynamics of a β-plume in the eastward stream can be obtained by considering the propagation of Rossby waves. The plume is formed by long Rossby waves propagating upstream of the eastward zonal current. The frequency of barotropic Rossby waves, $\omega$, is given by the dispersion relation

$$\omega = \bar{u}k - \frac{\beta k}{k^2 + l^2 + 1/R_b^2} \quad , \tag{17}$$

where $\bar{u}$ is mean zonal velocity, $k$ and $l$ are the wavenumbers in $x$ and $y$ directions, respectively and $R_b = (gh)^{1/2}/f_0$ is the barotropic radius of deformation. The $x$-component of the group velocity of zonally elongated ($k \approx 0$) Rossby waves is given by

$$c_{gx} = \bar{u} - \frac{\beta}{l^2 + 1/R_b^2} \quad , \tag{18}$$

while the y-component of the group velocity vanishes. Thus, when the meridional wavelength of the waves is large enough the waves are able to overcome the zonal flow and can travel westward. In our experiments, barotropic Rossby waves are emitted by baroclinic eddies in the area above the heater. Let us assume that the size of these eddies and hence the meridional wavelength of the emitted Rossby waves is determined by the baroclinic radius of deformation $R_c=(g'h_1)^{1/2}/f_0$ , where $g'$ is the reduced gravity and $h_1$ is the thickness of the upper layer. The critical value of the velocity of the zonal flow can then be estimated as

$$\bar{u} = \frac{\beta}{1/R_c^2 + 1/R_b^2} \, . \tag{19}$$

In order to estimate $R_c$, we calculated $g'$ using the empirical relation $g' \approx 25(B_0 r_h)^{2/3}/h$ given by Maxworthy and Narimousa[15] and took $h_1 = 0.5h$. Mean zonal velocity in (19) was taken to be $\bar{u} = U_{east}$. The solid line in Fig. 9 shows Eq. (19), which gives a reasonable estimate of the boundary between the regimes.

## IV. DISCUSSION AND CONCLUSIONS

In this work, we have shown that two distinct regimes exist in the flow generated by a convective source embedded in a zonal current. The main control parameters of the problem are the strength of the heater, which can be measured in terms of the upwelling velocity $w$ (which, in turn, is determined by the buoyancy flux per unit area, $B_0$ and the Coriolis parameter, $f_0$) and the strength of the sink (which, together with $\beta$, determines the velocity of the background zonal current). The flow induced by a heater without the background zonal flow is in the form of a β-plume. Zonal jets of alternating directions are formed within the plume. The (steady-state) westward transport velocity in the plume is proportional to the upwelling velocity above the heater in agreement with linear theory. When the background flow in the form of an eastward zonal current is present, the β-plume can be overwhelmed by the eastward current such that the warm water upwelled to the surface above the heater is advected to the east. The regime diagram shows the area where β-plume can exist in the parameter space. A consideration of barotropic Rossby waves emitted by the baroclinic eddies in the heated area allows us to obtain the critical value of the velocity of the zonal flow below which the β-plume can exist.

In view of the results on the regime of the flow obtained in this work, we can offer an explanation for why the convective storm on Saturn did not generate a westward propagating plume. Let us take the radius of the storm to be approximately $R_s = 2500$ km and assume that the radius of deformation is of similar magnitude $R_b = 2500$ km (see Read et al.[18]). Assuming further that the meridional wavelength of the Rossby wave emitted by the storm is equal to 4 $R_s$ we can estimate the group velocity of zonally elongated Rossby waves to be

$$c_{gx} = \frac{\beta}{(2\pi/4R)^2 + 1/R_b^2} \approx 9 \text{ m/s} \; .$$

This velocity is not enough to overcome strong eastward current on the northern and southern flanks of the storm, but it seems to be sufficient to provide some westward transport at the latitude of the center of the storm. However, Rossby waves of relatively small meridional extent, apart from the variation of the Coriolis parameter, are influenced by the variation of the relative vorticity of the background zonal flow. The effective beta parameter, $\beta_{\text{eff}}$ can be determined as a gradient of the absolute vorticity

$$\beta_{eff} = \beta - \partial^2 \overline{u} / \partial y^2 \, .$$

According to Aguilar et al.[19] the values of $\beta_{\text{eff}}$ are in fact negative at the axis of the westward jet where the center of the storm is located and become zero at the flanks of the jet. Thus, one can expect an eastward-propagating plume at this latitude rather than a “regular” westward-propagating one. Since the flow is dominated by the advection due to strong eastward jets, the effect (if any) of the eastward β-plume is hardly noticeable.

**ACKNOWLEDGMENTS**

YDA gratefully acknowledges the support of the Natural Sciences and Engineering Research Council of Canada. We would like to thank S. H. Curnoe for thorough proofreading of the draft of this paper.